\documentclass[twocolumn,showpacs,aps,prl,
groupedaddress,amssymb,amsmath,nobalancelastpage]{revtex4}

\usepackage{graphicx}
\usepackage{longtable}

\begin{document}


\title{Velocity Profiles in Slowly Sheared Bubble Rafts}

\author{John Lauridsen*}
\author{Gregory Chanan$^\dag$}
\author{Michael Dennin}
\affiliation{Department of Physics and Astronomy, University of
California at Irvine, Irvine, California 92697-4575}
\affiliation{*current address: Naval Surface Warfare Center Corona
Division QA-32, Corona,California 92878-5000} \affiliation{$^\dag$
current address: Stanford University, PO Box 11878 Stanford, CA
94309}

\date{\today}

\begin{abstract}

Measurements of average velocity profiles in a bubble raft
subjected to slow, steady-shear demonstrate the coexistence
between a flowing state and a jammed state similar to that
observed for three-dimensional foams and emulsions [Coussot {\it
et al,}, Phys. Rev. Lett. {\bf 88}, 218301 (2002)].  For
sufficiently slow shear, the flow is generated by nonlinear
topological rearrangements. We report on the connection between
this short-time motion of the bubbles and the long-time averages.
We find that velocity profiles for individual rearrangement events
fluctuate, but a smooth, average velocity is reached after
averaging over only a relatively few events.

\end{abstract}


\maketitle

A proposed jamming phase diagram \cite{LN98}, which treats applied
stress in a manner analogous to temperature and density, offers an
interesting framework for the study of a wide range of systems
subjected to shear \cite{LN01}. The jamming phase diagram proposes
the existence of a new state of matter, the ``jammed state'', and
suggests that similar transitions to this state would occur as a
function of applied stress, temperature, or density. For example,
many complex fluids, such as emulsions, foams, and granular
materials, exhibit a yield stress below which the material does
not flow, or jams. As one approaches the yield stress from above,
the viscosity diverges. This is analogous to the behavior of the
viscosity in the glass transition. Experiments have confirmed the
applicability of the jamming phase diagram in certain colloidal
systems \cite{TPCSW01}. These are examples of continuous jamming
transitions.

Recent work in a range of soft materials, including colloids,
granular matter, foams, and emulsions, suggest that the jamming
transition can be discontinuous \cite{CRBMGH02,DCBC02}. The
materials exhibited a yield stress. Yet, under conditions of
steady shear flow, there existed a critical radius at which the
shear-rate was discontinuous and a transition from a flowing to a
jammed state occurred \cite{CRBMGH02}. This is in contrast to the
results of most stress versus rate of strain measurements for
materials with a yield stress. Such experiments suggest the stress
is well-modelled as a continuous function of shear-rate, such as
with a Herschel-Bulkley model \cite{BAH77}. Additional evidence
for the discontinuous transition is given by the observation of a
bifurcation in the material's viscosity during constant stress
experiments \cite{DCBC02}.

In addition to the connection with the jamming transition, the
coexistence of a flowing and jammed state is one example of
another feature of some jammed systems: shear localization
\cite{HBV99,MDKENJ00,LBLG00,DTM01,BLSLG01,SCMM03}. In general,
shear localization (or banding) refers to a flow state which is
spatially separated into a regions of different shear rates. For
wormlike micelles, shear banding has been observed in which there
is a discontinuous transition from a high shear rate region to a
low shear rate region \cite{SCMM03}. Experiments in granular
systems \cite{HBV99,MDKENJ00,LBLG00,BLSLG01} and confined foam
\cite{DTM01} report a type of shear localization in which the
velocity profile is exponential. In these cases, the shear rate is
continuous, though the system is effectively divided into high
shear rate and zero shear rate regions. Simulations of
Lennard-Jones particles \cite{VBBB03} and quasi-static foam
\cite{KD03} confirm that shear localization can occur independent
of the existence of a yield stress. In the case of quasi-static
foam, the simulations suggest that a key element is the
localization of slip events \cite{KD03}. Understanding the
different sources of shear-localization will contribute to a
better understanding of the jamming phase diagram.

In this Letter, we confirm the coexistence between a flowing and a
jammed state for slow, steady shear of a bubble raft. By
considering the behavior on relatively short time scales, we show
the connection between fluctuations that occur in the stress and
the nonlinear flow events that comprise the unjammed state. This
provides insight into the connection between the individual bubble
motions and the observed average flow properties. The average
stress versus rate of strain curves have been measured separately
\cite{PD03}. These curves are consistent with a Herschel-Bulkley
model for the system. However, in light of the velocity
measurements reported here, it is important to reconsider the
interpretation of those results.

A bubble raft consists of a single layer of bubbles floating on a
fluid surface \cite{AK79}. Our bubble raft is described in detail
in Ref.~\cite{LTD02}. A random distribution of bubble sizes was
used, with an average radius of $1\ {\rm mm}$. The Couette
viscometer is described in detail in Ref.~\cite{app}. To shear the
foam, an outer Teflon barrier is rotated at a constant angular
velocity. We measure the azimuthal velocity, $v_{\theta}(r)$. The
shear rate is given by $\dot{\gamma}(r) =
r\frac{d}{dr}\frac{v_{\theta}(r)}{r}$ \cite{BAH77}. Therefore, we
considered the normalized angular velocity, $v(r) =
v_{\theta}(r)/(\Omega r)$, where the jammed state is $v(r) = 1$,
or $\dot{\gamma} = 0$. The stress, $\sigma(r_i)$, on the inner
cylinder was monitored by measuring the torque, $T$, on the inner
cylinder: $\sigma(r_i) = T/(2\pi r_i^2)$. The inner cylinder was
supported by a torsion wire (torsion constant $\kappa = 570\ {\rm
dyne\ cm}$), and $T$ was determined from the angular position of
the inner cylinder. Therefore, the inner cylinder had an
instantaneous angular speed, even though its average angular speed
is zero. At {\it both} boundaries, the first layer of bubbles was
never observed to slip relative to the boundary. This feature
combined with the finite size of bubbles set the effective inner
($r_i = 4.3\ {\rm cm}$) and outer ($R = 7.2\ {\rm cm}$) radii. We
report results for two rotation rates of the outer cylinder,
$\Omega = 8 \times 10^{-4}\ {\rm rad/s}$ and $\Omega = 5 \times
10^{-3}\ {\rm rad/s}$, which corresponds to shear rates (at $r =
4.3\ {\rm cm}$): $\dot{\gamma} = 4 \times 10^{-3}\ {\rm s^{-1}}$
and $\dot{\gamma} = 3 \times 10^{-2}\ {\rm s^{-1}}$.

The $<\sigma >$ versus $\dot{\gamma}$ behavior was measured
separately and is reported elsewhere \cite{PD03}. It is consistent
with a Herschel-Bulkley form for the viscosity: $<\sigma (r_i)> =
\mu_o \dot{\gamma}^n + \tau_o$, with $n = 1/3$, and $\tau_o = 0.8
\pm 0.1\ {\rm dyne/cm}$. For $\dot{\gamma} < 0.1\ {\rm s^{-1}}$,
the $<\sigma>$ was essentially independent of shear rate, though
it fluctuated between $0.5\ {\rm dyne/cm}$ and $2\ {\rm dyne/cm}$
\cite{PD03}. Both shear rates reported on here are within this
quasi-static limit. The gas area-fraction was 0.95. As the bubbles
are actually three-dimensional, we used an operational definition
of gas area-fraction: the ratio of area inside bubbles to the
total area of the bubble raft in a digitized image.

The fluid substrate (subphase) is driven at the same time as the
bubbles \cite{app}. Tests were made with bubble rafts that did not
touch the outer barrier. Under rotation of the outer barrier, no
motion of the bubbles was detected, ruling out driving by the
subphase. In addition, because the subphase is sheared with the
bubbles, the bubbles and subphase have similar velocities during
most of the motion. This minimizes any viscous dissipation between
bubbles and water. Also, the effective internal viscosity of the
bubble raft is 300 to 2500 g/cm~s for the range of $\dot{\gamma}$
studied. This is a factor of $10^4$ to $10^5$ greater than the
viscosity of the subphase. This ratio ensures that dissipation
between bubbles dominates dissipation from the subphase-bubble
interactions.

Bubble velocities were measured from taped video images that were
digitized in the computer. For $\Omega = 8 \times 10^{-4}\ {\rm
rad/s}$, the time between digitized images was 3.2 s/image, and
for $\Omega = 5 \times 10^{-3}\ {\rm rad/s}$, it was 2.0 s/image.
An image processing routine based on standard National Instruments
Labwindows$^{\rm TM}$/CVI functions detected and tracked
individual bubbles. The velocity was calculated using the total
displacement of 10 digitized images. This allowed tracking of the
rapidly moving bubbles from image to image (short time between
individual images) and sufficient total time to measure bubbles as
slow as $6 \times 10^{-4}\ {\rm cm/s}$ for $\Omega = 8 \times
10^{-4}\ {\rm rad/s}$ and $9 \times 10^{-4}\ {\rm cm/s}$ for
$\Omega = 5 \times 10^{-3}\ {\rm rad/s}$. Finally, the bubbles
within each of 24 equally spaced radial bands in the range $4.3
\le r \le 7.2 \ {\rm cm}$ were averaged to compute $v(r)$.

\begin{figure}
\includegraphics[width=8.4cm]{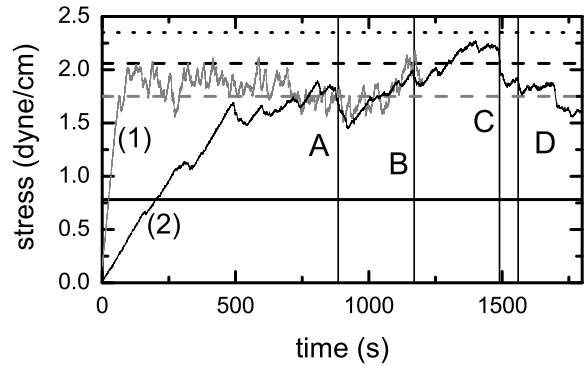}
\caption{\label{stress} Stress versus time for $\Omega = 5 \times
10^{-4}\ {\rm rad/s}$ [gray line labelled (1)] and $\Omega = 8
\times 10^{-4}\ {\rm rad/s}$ [black line labelled (2)]. The solid
horizontal line is $\tau_o$ determined from a fit using data from
Ref.~\cite{PD03} to a Herschel-Bulkley fluid model. The dashed
horizontal line is $\sigma(r_i)$ when $\sigma(6.7\ {\rm cm}) =
\tau_o$ (black line) or $\sigma(6.3\ {\rm cm}) = \tau_o$ (gray
line). The dotted line is $\sigma(r_i)$ when $\sigma(R) = \tau_o$.
The vertical lines indicate the stress drops presented in
Fig.~\ref{image} for $\Omega = 8 \times 10^{-4}\ {\rm rad/s}$.}
\end{figure}

\begin{figure}
\includegraphics[width=8.4cm]{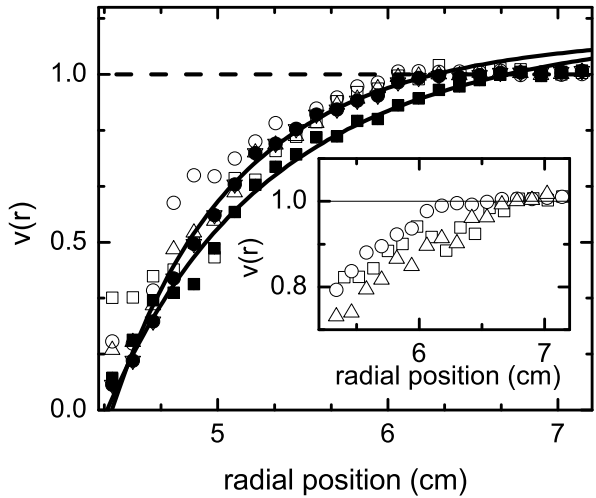}
\caption{\label{vel} $v(r) = v_{\theta}(r)/(r \Omega)$ versus $r$
for $\Omega = 5 \times 10^{-3}\ {\rm rad/s}$ averaged over time
from $t = 250\ {\rm s}$ to the end of the run ($\bullet$) and for
$\Omega = 8 \times 10^{-4}\ {\rm rad/s}$ averaged over time from
$t = 650\ {\rm s}$ to the end of the run ($\blacksquare$). The
solid lines are fits to a power-law model for viscosity. The
dashed line is the $v = 1$ line. Also shown are $v(r)$ for $\Omega
= 5 \times 10^{-3}\ {\rm rad/s}$ averaged over a single event
($\square$), four events ($\circ$), 10 events ($\vartriangle$),
and 20 events ($\triangledown$). The insert illustrates the
discontinuity in the shear rate for $\Omega = 5 \times 10^{-3}\
{\rm rad/s}$ ($\bullet$), $\Omega = 8 \times 10^{-4}\ {\rm rad/s}$
($\vartriangle$), and $\Omega = 1 \times 10^{-4}\ {\rm rad/s}$
($\square$).}
\end{figure}

Figure~\ref{stress} is a plot of $\sigma(r_i)$ versus time for the
velocity data reported on here. The initial elastic regime
consists of a linear increase of $\sigma(r_i)$ with time. The
subsequent ``flowing'' regime is dominated by irregular variations
in the stress characteristic of the slow shear-rate ``flow'' of
many jammed systems \cite{LN01}. For comparison with the flow data
presented later, the solid line in Fig.~\ref{stress} represents
$\tau_o$ (the ``yield stress'') based on the fit of $<\sigma
(r_i)>$ to a Herschel-Bulkley model \cite{PD03}. (Given the
existence of a discontinuous transition, such a definition of a
yield stress may not be meaningful.) For each data set, the dashed
line represents the value of $\sigma(r_i)$ such that $\sigma(r_c)
= \tau_o$, assuming $\sigma (r) = T/(2\pi r^2)$ \cite{BAH77}.
($r_c$ is the radius at which the system jams, see
Fig.~\ref{vel}). The dotted line is $\sigma(r_i)$ such that
$\sigma(R) = \tau_o$.

Figure~\ref{vel} shows $v(r)$ versus $r$ for $\Omega = 5 \times
10^{-3}\ {\rm rad/s}$ for a number of different averages. In each
case, the average is computed using bubbles from roughly $1/3$ of
the system. The solid circles represent an average over $1000\
{\rm s}$, starting $210\ {\rm s}$ after the initiation of shear.
This corresponds to approximately 2800 individual velocities per
channel. Defining an event as a consecutive period of stress
increase and decrease, the other curves are averaged over a single
event ($\square$), four events ($\circ$), 10 events
($\vartriangle$), and 20 events ($\triangledown$), respectively.
The 10 event average is in reasonable agreement with $1000\ {\rm
s}$ average, and the 20 event average (400~s) is indistinguishable
from the $1000\ {\rm s}$ average. Figure~\ref{vel} also shows
$v(r)$ versus $r$ for $\Omega = 8 \times 10^{-4}\ {\rm rad/s}$.
The average covered a total time of $1020\ {\rm s}$, starting
$650\ {\rm s}$ after the initiation of shear. This corresponds to
approximately 2000 individual velocities per channel and on the
order of 50 events.

To find $\dot{\gamma}$ and $r_c$, the average velocity is fit to
$v(r) = A + B/r^{2/n}$ (solid curves) over the range $0 < v(r) <
0.98$. This is the velocity profile for a power-law fluid
($\sigma(r) \propto r^n$) in a Couette geometry \cite{BAH77}. A
number of aspects of this fit differ from the expected solution
for a Herschel-Bulkley fluid. First, there is a discontinuity in
$\dot{\gamma}$. This is highlighted by the insert in
Fig.~\ref{vel}. Here data is shown for three rotation rates,
$\Omega = 5 \times 10^{-3}\ {\rm rad/s}$, $\Omega = 8 \times
10^{-4}\ {\rm rad/s}$, and $\Omega = 1 \times 10^{-4}\ {\rm
rad/s}$. (For $\Omega = 1 \times 10^{-4}\ {\rm rad/s}$, there are
only approximately 10 events. Based on the results for $\Omega = 5
\times 10^{-3}\ {\rm rad/s}$, this is sufficient to confirm the
discontinuity.) Quantitatively, the crossing of the fit to a
power-law velocity profile and $v(r) = 1$ defines the critical
radius, $r_c$, and critical shear rate, $\dot{\gamma}(r_c)$. For
$\Omega = 8 \times 10^{-4}\ {\rm rad/s}$, $r_c = 6.7\ {\rm cm}$,
and $\dot{\gamma}(r_c) = 6 \times 10^{-4}\ {\rm s^{-1}}$. For
$\Omega = 5 \times 10^{-3}\ {\rm rad/s}$, $r_c = 6.3\ {\rm cm}$,
and $\dot{\gamma}(r_c) = 4 \times 10^{-3}\ {\rm s^{-1}}$.  Unlike
in some foam experiments \cite{RBC03}, the critical shear rate
differs for the two speeds. Second, the fits give $n = 0.45 \pm
0.05$ for $\Omega = 8 \times 10^{-4}\ {\rm rad/s}$ and $n = 0.33
\pm 0.02$ for $\Omega = 5 \times 10^{-3}\ {\rm rad/s}$. For the
faster rotation rate, the exponent is in agreement with the
exponent in the Herschel-Bulkley fit to the stress \cite{PD03} and
velocity profiles measured at higher shear rates \cite{LTD02}.
However, the measured exponent is different for the two rotation
rates, as seen in other systems \cite{CRBMGH02}.

The trend for average $r_c$ is opposite the expectation for a
Herschel-Bulkley fluid, where a continuous decrease in $<\sigma>$
results in a continuous decrease in $r_c$. It is consistent with
the measured behavior of $<\sigma (r_i)>$ in the quasi-static
regime. For $\Omega = 8 \times 10^{-4}\ {\rm rad/s}$, the system
spends more time at a higher value of stress (see
Fig.~\ref{stress}), and $r_c$ is larger. This is also different
then trends reported in Ref.~\cite{CRBMGH02} and \cite{RBC03}. for
higher shear-rates in the ``continuum'' regime. This difference is
not surprising given that the behavior in the continuum regime is
``smoother'' than in the quasi-static, or ``discrete'', regime in
which we worked.

\begin{figure}
\includegraphics[width=8.4cm]{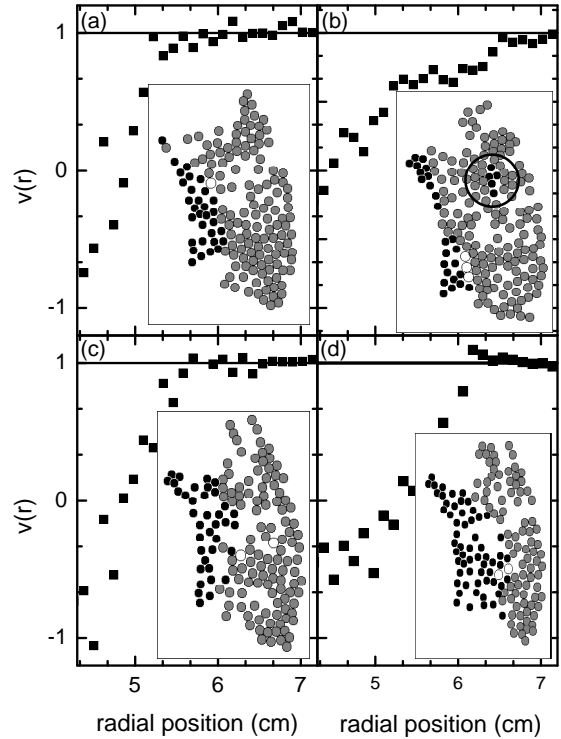}
\caption{\label{image} Average velocity during the stress drops
indicated in Fig.~\ref{stress} for $\Omega = 8 \times 10^{-4}\
{\rm rad/s}$. The inserts are images of a subset of the bubbles,
with gray circles corresponding to $v(r)$ in the same sense as
$\Omega$, black circles corresponding to $v(r)$ opposite $\Omega$,
and white circles $|v(r)| \leq 6 \times 10^{-4}\ {\rm cm/s}$.}
\end{figure}

Figure~\ref{image} highlights the importance of considering
short-time velocity profiles, in addition to the average
quantities. In Fig.~\ref{image}, plots of velocity profiles
averaged over a single stress drop and corresponding snapshots of
the bubble motions are shown. The stress drops corresponding to
Figs.~\ref{image}(a)-(d) are indicated by vertical lines in
Fig.~\ref{stress}, labelled with corresponding letters. Here, the
shear-rate discontinuity is more apparent. The individual velocity
profiles are highly nonlinear and not consistent with a simple
continuum model for viscosity. As expected $r_c$ (the transition
point to elastic flow) fluctuates. However, these fluctuations are
not consistent with the continuum expectation of $\sigma \propto
1/r^2$ that predicts $r_c = [T/(2\pi \tau_o)]^{1/2}$, with $T =
2\pi r_i^2\sigma(r_i)$ \cite{BAH77}. For such a model, stress drop
(C) would have the largest value of $r_c$ given its value of
$\sigma(r_i)$. However, $r_c$ is greater for both (D) and (B).
This behavior is indicative of stress chains, or other nonuniform
stress distributions, existing in the foam, similar to those
observed for granular disks in two-dimensions \cite{HBV99}.

Snapshots of the selected bubble motions are presented as inserts
in Fig.~\ref{image}. The images are color coded so that bubbles
moving opposite the outer cylinder are black, gray bubbles are
moving in the direction of rotation, and white bubbles have
$|v(r)| \leq 6 \times 10^{-4}\ {\rm cm/s}$. The row of bubbles at
each boundary is not shown. The snapshots confirm that $r_c$ is
connected to bubble rearrangements. Both (B) and (D) have the
largest $r_c$ and radial position at which negative velocities are
observed (see circled region in Fig.~\ref{image}b).

In summary, we have observed the coexistence of a jammed and a
flowing state in a bubble raft in the quasi-static shear limit, as
has been observed for other soft matter systems at higher
shear-rates \cite{CRBMGH02}. For the average velocity profiles,
the transition appears to be discontinuous and occurs at a
critical radius set by the yield stress. These are two ways that
the average profiles for the bubble raft differ from observed
exponential shear-localization in confined, two-dimensional foam
\cite{DTM01} and granular systems
\cite{HBV99,MDKENJ00,LBLG00,BLSLG01}. For these systems, the
critical radius is significantly smaller and the shear-rate is
continuous. At this point, the reasons for the differences are not
clear, though one potential candidate is the role of viscous
dissipation. For the confined foam, simulations suggest that the
exponential velocity profile observed is due to localization of
the nonlinear rearrangements \cite{KD03}. These simulations do not
include viscous dissipation \cite{KD03}. Because this localization
is not observed in the bubble raft (see Fig.~\ref{image},
especially event C and D), there may exist differences in the role
of viscous dissipation in the bubble raft and the confined foam.
Another potential difference is the yield {\it strain}, which is
quite large in the bubble raft system \cite{LTD02}. This could
impact the distribution of rearrangement events. The results of
Fig.~\ref{image} point to the need to understand the velocity
profiles during individual events in order to resolve these
outstanding issues regarding the average behavior.

The results presented in Fig.~\ref{image} raise two important
questions. First, given the highly nonlinear and fluctuating
character of the velocities during individual events (see
Fig.~\ref{image}), why does the average velocity converge to a
smooth curve after averaging over only 20 such events (and even
fewer)? Second, what sets the critical radius for the individual
events (Fig.~\ref{image}) and how is the distribution of $r_c$ for
these events related to the value of $r_c$ found for the long-time
averages (Fig.~\ref{vel})? The determination of the critical
radius is particularly interesting given the results presented in
Fig.~\ref{image} and the work in other systems. Unlike the work in
the continuum limit \cite{CRBMGH02}, $r_c$ does not appear to be
set by a critical shear rate either for the average profiles (see
Fig.~\ref{vel}) or the short-time profiles. In fact, the concept
of shear rate is not well-defined for the motion during individual
events, and yet each event has a well-defined value of $r_c$. For
individual events, more work is needed to understand the
connection between stress and $r_c$. However, $r_c$ is clearly not
set by the stress on the inner boundary, suggesting the need to
understand the stress distribution within the bubble raft.
Finally, understanding the connection between the individual and
average quantities is necessary to resolve the apparent
discrepancy between $<\sigma>$ versus $\dot{\gamma}$ measurements
that suggest a continuous transition and the discontinuous
velocity profiles presented here. This is necessary for deeper
insight into the implications of the jamming phase diagram for
slow, steady-shear.

\begin{acknowledgments}

This work was supported by Department of Energy grant
DE-FG02-03ED46071, the Research Corporation and Alfred P. Sloan
Foundation. J. Lauridsen thanks UROP for additional funding for
this work. The authors thank Corey O'Hern and Phillipe Coussot for
useful discussions.

\end{acknowledgments}

\end{document}